\documentstyle[12pt]{article}
\begin{document}
\baselineskip=12pt
\pagestyle{empty}
\begin{flushright}
MUTP/29/93
\end{flushright}
\today
\bigskip
\bigskip
\baselineskip=24pt
\begin{center}
\begin{large}
{\bf Diffeomorphisms of the
Klein Bottle and Projective Plane}\\
\end{large}
\bigskip
\medskip
{\em J.S. Apps and J.S. Dowker} \\
\medskip
Department of Theoretical Physics \\
The University, Manchester M13 9PL \\
\bigskip
\medskip
{\bf ABSTRACT }
\end{center}
We calculate the Riemann curvature tensor and sectional curvature for the Lie
group of volume-preserving diffeomorphisms of the Klein bottle and projective
plane. In particular,
we investigate the sign of the sectional curvature, and find a possible
disagreement with a theorem of Lukatskii. We suggest an amendment to
this theorem.
\newpage
\pagestyle{plain}
\pagenumbering{arabic}
\vspace{1in}

\section{Introduction}

The theory of diffeomorphisms on two-dimensional manifolds can be applied to
ideal fluid dynamics, as well as to relativistic membranes. In the
former case, we can see a diffeomorphism on a two-manifold $\cal M$ as an
infinitely differentiable mapping of one fluid configuration onto another. The
set of possible diffeomorphisms forms an infinite-dimensional Lie group
manifold, denoted by Diff$\cal M$, upon which we choose the identity to be an
arbitrary configuration of the fluid. The fluid flow is then given by a curve
on Diff$\cal M$ which is parameterized by time.
 In particular, diffeomorphisms which preserve the volume element on $\cal M$
form a Lie subgroup of Diff$\cal M$, denoted by SDiff$\cal M$. These are
related to the flow of an incompressible fluid.

With the application of a least-action principle for the fluid, the curve on
SDiff$\cal M$ can be shown to be a geodesic \cite {ebin}. As the curvature of
the space affects the deviation of nearby geodesics, calculation of the
curvature tensor for SDiff$\cal M$ will give information about the behaviour of
the fluid on $\cal M$. More precisely, a negative sectional curvature will
cause
geodesics to diverge, while if it is positive, they will remain in the
same neighbourhood. In this way, it is possible to predict whether or not two
initially similar fluid flows will remain similar, i.e. whether a flow is
``stable''.

 Arnold's fundamental paper
\cite {arnold} calculates the curvature of the volume-preserving diffeomorphism
group for the torus, SDiff$T^{2}$. The calculation has been repeated for the
two-sphere by Lukatskii \cite {lukspher} (see also \cite {dandwei}), for the
three-sphere by
Dowker \cite {dowker} and for the flat Riemann surfaces by Wolski and Dowker
\cite {wolskitet}, \cite {wolskigen}. In this paper, we apply Arnold's
calculation to the group of volume-preserving diffeomorphisms of the Klein
bottle and projective plane.

\section {Formalism}

The generators of volume-preserving diffeomorphisms on a locally Euclidean
two-manifold $\cal M$ have the form
\begin {equation}
\label {generatordefn}
v_{k}=\epsilon^{ab}\partial_{b}\phi_{k}\:\partial_{a}
\end {equation}
where $\phi_{k}$ is some scalar function on $\cal M$. A complete set of
diffeomorphisms on $\cal M$ given by a complete set of functions $\phi_{k}$ is
a Lie algebra, i.e.
\begin {equation}
\label {liebracketdefn}
[v_{k},v_{l}]=\sum_{m}f^{m}_{\:\:\:\:kl}L_{m}
\end {equation}
where the $f^{m}_{\:\:\:\:kl}$ are the structure constants.
 This is the Lie algebra of the infinite-dimensional Lie group SDiff$\cal M$.

The kinetic energy of the fluid on $\cal M$ is used to define the metric at the
identity of SDiff$\cal M$.
\begin {eqnarray}
\label {metricdefn}
\langle v_{k},v_{l} \rangle &=&\int_{\cal M}d^{2}{\bf
x}\:(\epsilon^{ab}\partial
_{b}\phi_{k})(\epsilon^{cd}\partial_{d}\phi_{l})\:\partial_{a}\cdot
\partial_{c}\\
\label {metricform}
&=&\int_{\cal M}d^{2}{\bf x}\:(\nabla \phi_{k})\cdot (\nabla \phi_{l})
\end {eqnarray}
since the metric on $\cal M$ is Euclidean:
\begin {equation}
\label {localmetric}
\partial_{a}\cdot \partial_{c}=\delta_{ac}
\end {equation}
The kinetic energy
 is invariant under the right action of the diffeomorphism group,
since it is independent of the fluid configuration.

With the intention of calculating the curvature of SDiff$\cal M$, we need an
expression for the covariant derivative. We follow the approach of Milnor
\cite {milnor} (though in our own notation).
 If the metric is right-invariant, then for
right-invariant fields $\tilde{v}_{k}$, equal to $v_{k}$ at the identity, we
have
\begin {equation}
\label {constantmetric}
\nabla_{\tilde{v}_{k}}\langle \tilde{v}_{l},\tilde{v}_{m} \rangle =0
\end {equation}
so that
\begin {equation}
\label {fundamental}
\langle \nabla_{\tilde{v}_{k}}\tilde{v}_{l},\tilde{v}_{m} \rangle +\langle
\tilde{v}_{l},\nabla_{\tilde{v}_{k}} \tilde{v}_{m}\rangle =0
\end {equation}
{}From now on we drop the tilde and designate the right-invariant fields on
SDiff$\cal M$ by $v_{k}$. If we use the torsionless definition of the covariant
derivative:
\begin {equation}
\label {torsionless}
[v_{k},v_{l}]=\nabla_{v_{k}}v_{l}-\nabla_{v_{l}}v_{k}
\end {equation}
then from (\ref {fundamental}) we can derive, by permuting $k$, $l$ and $m$:
\begin {equation}
2\langle v_{k},\nabla_{v_{l}}v_{m}\rangle=\langle v_{k},[v_{l},v_{m}] \rangle
+\langle v_{m},[v_{k},v_{l}]\rangle -\langle v_{l},[v_{m},v_{k}]\rangle
\end {equation}
To find an explicit expression for $\nabla_{v_{l}}v_{m}$, Arnold \cite {arnold}
defines the vectors $B(v_{k},v_{l})$ such that
\begin {equation}
\label {bdefn}
\langle v_{k},[v_{l},v_{m}] \rangle =\langle v_{m},B(v_{k},v_{l})\rangle
\end {equation}
We then obtain
\begin {equation}
\label {grad}
2\nabla_{v_{l}}v_{m}=[v_{l},v_{m}]-B(v_{l},v_{m})-B(v_{m},v_{l})
\end {equation}
due to the completeness of the set $\{v_{k}\}$. The $B$ vectors can be found by
expanding on the basis $\{v_{k}\}$:
\begin {equation}
\label {bexp}
B(v_{k},v_{l})=\sum_{m}b^{m}_{\:\:\:\:kl}v_{m}
\end {equation}
and then using the definition (\ref {bdefn}) to calculate the
$b^{m}_{\:\:\:\:kl}$.
Note that (\ref {grad}) holds only for right-invariant fields.

As we have, from (\ref {liebracketdefn}) and (\ref {metricform}), the Lie
bracket and metric at the identity, we can therefore calculate the covariant
derivative at the identity. The realization that, for right-invariant fields,
the structure constants in (\ref {liebracketdefn}) are the same at every point
on the Lie group manifold, then allows us to calculate the second derivative,
since the $f^{m}_{\:\:\:\:kl}$ are constant
with respect to the covariant derivative.

The Riemann curvature tensor is defined by
\begin {equation}
\label {riemanndefn}
{\bf R}(v_{k},v_{l})v_{m}=\nabla_{[v_{k},v_{l}]}v_{m}-
[\nabla_{v_{k}},\nabla_{v_{l}}]v_{m}
\end {equation}
and its covariant components are
\begin {equation}
\label {riemanncomp}
R_{v_{k},v_{l},v_{m},v_{n}}=\langle
\nabla_{[v_{k},v_{l}]}v_{m}-[\nabla_{v_{k}},\nabla_{v_{l}}]v_{m},v_{n}\rangle
\end {equation}

Additionally, in this paper we evaluate the normalized sectional curvature, and
  use the general definition:
\begin {equation}
\label {sectionaldefn}
C_{v_{k},v_{l}}=\frac{R_{v_{k},v_{l},v_{k},v_{l}}}
{\langle v_{k},v_{k} \rangle \langle v_{l},v_{l}
 \rangle -\langle v_{k},v_{l} \rangle ^{2}}
\end {equation}
The sign of the above quantity is important, since it affects the stability of
the fluid flow generated by $v_{k}$ and $v_{l}$. Lukatskii \cite {lukatskii}
has published a theorem which states that on a locally Euclidean manifold,
 the sectional curvature defined by
any two flows, of which at least one is {\bf stationary}, is non-positive,
where a stationary flow is one for which the fluid velocity field is
time-independent. It can be shown (e.g. \cite {nakamura})
 that the fluid velocity satisfies
\begin {equation}
\label {geo}
\frac{\partial v}{\partial t}=-\nabla_{v}v
\end {equation}
where our definition of $\nabla_{v}v$ is different to that in
\cite {nakamura} as we
have defined $\nabla$ to be a vector on SDiff$\cal M$. Using (\ref {grad})
leads to the equation for vectors $v$ which generate stationary flows:
\begin {equation}
\label {stat}
B(v,v)=0
\end {equation}

\section {The Curvature of SDiff$T^{2}$}

In calculating the curvature of SDiff(Klein bottle), we use many of Arnold's
results \cite {arnold} for the torus. In this section, we quote these
results.

Arnold considers a $2\pi \times 2\pi$ torus, constructed by making the
identifications
\begin {eqnarray}
\label {torusxident}
(x,y)&=&(x+2\pi ,y)\\
\label {torusyident}
(x,y)&=&(x,y+2\pi )
\end {eqnarray}
on a plane. The basis functions are chosen to be the eigenfunctions of the
laplacian, i.e.
\begin {equation}
\label {waves}
\phi _{k}=e^{i{\bf k}\cdot {\bf x}}
\end {equation}
where ${\bf k}$ is a two-vector
\begin {equation}
{\bf k}=\left( \begin {array}{c} k_{1}\\k_{2}\end {array}\right)
\end {equation}
and $k_{1}$ and $k_{2}$ are integers, so that the $\phi_{k}$ are invariant
under (\ref {torusxident}) and (\ref {torusyident}). The diffeomorphism
generators are therefore
\begin {equation}
\label {torusgenerator}
L_{k}=\epsilon^{ab}\partial_{b}e^{i{\bf k}\cdot {\bf x}}\:\partial_{a}
\end {equation}
and, from (\ref {metricform}), we have the scalar product
\begin {equation}
\label {torusmetric}
\langle L_{k},L_{l} \rangle =4\pi ^{2}{\bf k}^{2}\delta_{{\bf k}+{\bf l}}
\end {equation}
where we define $\delta_{a}\equiv \delta_{a,0}$

The Lie bracket is simply the commutator of the derivatives (\ref
{torusgenerator}):
\begin {eqnarray}
[L_{k},L_{l}]&=&\epsilon^{ab}\partial_{b}\phi_{k}\:\partial_{a}(\epsilon^{cd}
\partial_{d}\phi_{l}\:\partial_{c})
-\epsilon^{ab}\partial_{b}\phi_{l}\:\partial_{a}
(\epsilon^{cd}\partial_{d}\phi_{k}\:\partial_{c})\\
\label {torusliebracket}
&=&({\bf k}\times {\bf l})L_{k+l}
\end {eqnarray}
where ${\bf k}\times {\bf l}$ is the 2-dimensional cross product
$k_{1}l_{2}-k_{2}l_{1}$. Inserting (\ref {torusmetric}) and (\ref
{torusliebracket}) in (\ref {bdefn}) we obtain the expansion coefficients for
$B(L_{k},L_{l})$:
\begin {equation}
b^{m}_{\:\:\:\:kl}=({\bf m}\times {\bf l})\frac{{\bf k}^{2}}{{\bf m}^{2}}\delta
_{{\bf k}+{\bf l}-{\bf m}}
\end {equation}
so that
\begin {equation}
\label {torusb}
B(L_{k},L_{l})=({\bf k}\times {\bf l})\frac {{\bf k}^{2}}{({\bf k}+{\bf
l})^{2}}
L_{k+l}
\end {equation}

The calculation of the Riemann tensor is now straightforward. We simplify the
notation by defining, for use in the next section:
\begin {eqnarray}
f_{k,l}&=&{\bf k}\times {\bf l}\\
g_{k,l}&=&\frac{{\bf k}^{2}}{({\bf k}+{\bf l})^{2}}({\bf k}\times {\bf l})\\
h_{k,l}&=&\frac{1}{2}(f_{k,l}-g_{k,l}-g_{l,k})\\
t_{k,l,m}&=&f_{k,l}h_{k+l,m}-h_{l,m}h_{k,l+m}+h_{k,m}h_{l,k+m}
\end {eqnarray}

Arnold's results can then be summarized:
\begin {eqnarray}
\nabla_{L_{k}}L_{l}&=&h_{k,l}L_{k+l}\\
\nabla_{L_{k}}\nabla_{L_{l}}L_{m}&=&h_{l,m}h_{k,l+m}L_{k,l+m}\\
{\bf R}(L_{k},L_{l})L_{m}&=&t_{k,l,m}L_{k+l+m}
\end {eqnarray}
where we have used the definition (\ref {riemanndefn}). The components of the
Riemann tensor become
\begin {equation}
R_{L_{k},L_{l},L_{m},L_{n}}=
4\pi ^{2}{\bf n}^{2}t_{k,l,m}\delta_{{\bf k}+{\bf l}+{\bf m}+{\bf n}}
\end {equation}
Arnold shows (though we will not go into detail here --
the proof simply involves
two-vector algebra) that the components can be written
\begin {equation}
\label {theformula}
R_{L_{k},L_{l},L_{m},L_{n}}=
4\pi ^{2}\left[\frac{({\bf k}\times {\bf m})^{2}({\bf l}\times {\bf
n})^{2}}{|{\bf k}+{\bf m}||{\bf l}+{\bf n}|}-\frac{({\bf k}\times {\bf n})^{2}
({\bf l}\times {\bf m})^{2}}{|{\bf k}+{\bf n}||{\bf l}+{\bf m}|}\right]
\delta_{{\bf k}+{\bf l}+{\bf m}+{\bf n}}
\end {equation}

\section {The Curvature of SDiff(Klein Bottle)}

The Klein bottle can be seen as a M\"{o}bius band with periodic boundary
conditions. A $2\pi \times \pi$ Klein bottle can be constructed from the $2\pi
\times 2\pi$ plane by making the identifications
\begin {eqnarray}
\label {bottlexident}
(x,y)&=&(x+2\pi ,y)\\
\label {bottleyident}
(x,y)&=&(2\pi -x,y+\pi)
\end {eqnarray}

Due to the reflection (\ref {bottleyident}), this is a non-orientable manifold,
although it has the same intrinsic geometry as the torus. Any diffeomorphism on
the manifold must be invariant under the above identifications -- this places a
restriction on the possible diffeomorphism generators. As Pope and Romans \cite
{pope} have stated, the allowed generators are
\begin {equation}
\label {bottlegenerator}
e_{k}=L_{k}-(-)^{k_{2}}L_{\overline {k}}
\end {equation}
where
\begin {equation}
\overline {{\bf k}}=\left( \begin {array}{c}-k_{1}\\k_{2}\end {array}
\right)
\end {equation}
and the $L_{k}$ are the diffeomorphism generators on the torus. The $e_{k}$
have the required properties
\begin {eqnarray}
e_{k}(x,y)&=&e_{k}(x+2\pi ,y)\\
e_{k}(x,y)&=&e_{k}(2\pi -x,y+\pi )
\end {eqnarray}
and are hence invariant under (\ref {bottlexident}) and (\ref {bottleyident}).
Note that (\ref {bottlegenerator}) also generate a subgroup of diffeomorphisms
on the torus.

Although there is no continuous volume-form on the Klein bottle, there is no
problem with integrating a scalar field over this manifold, and we can obtain
the metric in the same way as for the torus. In the fluid dynamical
interpretation, this is just the kinetic energy of the fluid on the Klein
bottle, which is well-defined.
 The metric at the
identity on SDiff(Klein bottle) is
\begin {eqnarray}
\langle e_{k},e_{l}\rangle &=&\int_{x=0}^{2\pi}\int_{y=0}^{\pi}dx\:dy\:\nabla
[e^{i{\bf k}\cdot {\bf x}}-(-)^{k_{2}}e^{i\overline {{\bf k}}\cdot {\bf
x}}]\cdot \nabla [e^{i{\bf l}\cdot {\bf x}}-(-)^{l_{2}}e^{i\overline
 {{\bf l}}\cdot {\bf x}}]\\
&=&\frac{1}{2}\int_{x=0}^{2\pi}\int_{y=0}^{2\pi}dx\:dy\:\nabla
[e^{i{\bf k}\cdot {\bf x}}-(-)^{k_{2}}e^{i\overline {{\bf k}}\cdot {\bf
x}}]\cdot \nabla [e^{i{\bf l}\cdot {\bf x}}-(-)^{l_{2}}e^{i\overline
 {{\bf l}}\cdot {\bf x}}]\\
\label {bottlemetric}
&=&4\pi ^{2}{\bf k}^{2}[\delta_{{\bf k}+{\bf l}}-(-)^{k_{2}}\delta_{\overline
{{\bf k}}+{\bf l}}]
\end {eqnarray}

The Lie bracket can be derived using linearity:
\begin {eqnarray}
[e_{k},e_{l}]&=&[L_{k},L_{l}]-(-)^{k_{2}}[L_{\overline {k}},L_{l}]-(-)^{l_{2}}
[L_{k},L_{\overline {l}}]+(-)^{k_{2}+l_{2}}
[L_{\overline {k}},L_{\overline {l}}]\\
\label {bottleliebracket}
&=&({\bf k}\times {\bf l})e_{k+l}-(-)^{k_{2}}(\overline {{\bf k}}\times {\bf
l})e_{\overline {k}+l}
\end {eqnarray}

To calculate $B(e_{k},e_{l})$, we expand on the set $\{e_{k}\}$. From
(\ref {bottlegenerator}), we see that
\begin {equation}
e_{\overline {k}}=-(-)^{k_{2}}e_{k}
\end {equation}
so that $e_{k}=0$ for $k_{1}=0$, $k_{2}$ even, and that there is no need to sum
over negative values of $k_{1}$ in an expansion. Alternatively, we can sum over
all values of $k_{1}$ and $k_{2}$:
\begin {equation}
B(e_{k},e_{l})=\sum_{m_{1}=-\infty}^{\infty}\sum_{m_{2}=-\infty}^{\infty}
b^{m}_{\:\:\:\:kl}e_{m}
\end {equation}
and then impose the condition
\begin {equation}
\label {bcond}
b^{m}_{\:\:\:\:kl}=-(-)^{m_{2}}b^{\overline {m}}_{\:\:\:\:kl}
\end {equation}
without losing any generality. Using (\ref {bdefn}) together with (\ref
{bcond}), we arrive at
\begin {equation}
\label {bottleb}
B(e_{k},e_{l})={\bf k}^{2}\left[ \frac{({\bf k}\times {\bf l})}{({\bf k}+{\bf
l})^{2}}e_{k+l}-(-)^{k_{2}}\frac{(\overline {{\bf k}}\times {\bf
l})}{(\overline{{\bf k}}+{\bf l})^{2}}e_{\overline {k}+l}\right]
\end {equation}

Writing these results in terms of the notation of section 3, we have
\begin {eqnarray}
[e_{k},e_{l}]&=&f_{k,l}e_{k+l}-(-)^{k_{2}}f_{\overline {k},l}e_{\overline
{k}+l}\\
B(e_{k},e_{l})&=&g_{k,l}e_{k+l}-(-)^{k_{2}}g_{\overline {k},l}e_{\overline
{k}+l}
\end {eqnarray}

It follows that
\begin {eqnarray}
\nabla_{e_{k}}e_{l}&=&h_{k,l}e_{k+l}-(-)^{k_{2}}h_{\overline{k},l}
e_{\overline{k}+l}\\
\nabla_{e_{k}}\nabla_{e_{l}}e_{m}&=&h_{l,m}h_{k,l+m}e_{k+l+m}
-(-)^{k_{2}}h_{l,m}h_{\overline{k},l+m}e_{\overline{k}+l+m}\nonumber \\
& &-(-)^{l_{2}}h_{\overline{l},m}h_{k,\overline{l}+m}e_{k+\overline{l}+m}
+(-)^{k_{2}+l_{2}}h_{\overline{l},m}h_{\overline{k},\overline{l}+m}
e_{\overline{k}+\overline{l}+m}\\
\nabla_{[e_{k},e_{l}]}e_{m}&=&f_{k,l}h_{k+l,m}e_{k+l+m}
-(-)^{k_{2}}f_{\overline{k},l}h_{\overline{k}+l,m}e_{\overline{k}+l+m}\nonumber
\\& &-(-)^{l_{2}}f_{k,\overline{l}}h_{k+\overline{l},m}e_{k+\overline{l}+m}
+(-)^{k_{2}+l_{2}}f_{\overline{k},\overline{l}}h_{\overline{k}+\overline{l},m}
e_{\overline{k}+\overline{l}+m}
\end {eqnarray}

The Riemann tensor for the group of volume-preserving diffeomorphisms of the
Klein bottle is therefore
\begin {eqnarray}
{\bf R}(e_{k},e_{l})e_{m}&=&t_{k,l,m}e_{k+l+m}-(-)^{k_{2}}t_{\overline{k},l,m}
e_{\overline{k}+l+m}\nonumber \\
& &-(-)^{l_{2}}t_{k,\overline{l},m}e_{k+\overline{l}+m}
+(-)^{k_{2}+l_{2}}t_{\overline{k},\overline{l},m}
e_{\overline{k}+\overline{l}+m}
\end {eqnarray}
Its components become, using the metric (\ref {bottlemetric}):
\begin {eqnarray}
\label {result}
R_{e_{k},e_{l},e_{m},e_{n}}&=&T_{k,l,m,n}-(-)^{k_{2}}T_{\overline{k},l,m,n}
-(-)^{l_{2}}T_{k,\overline{l},m,n}-(-)^{m_{2}}T_{k,l,\overline{m},n}\nonumber
\\
& &+(-)^{k_{2}+l_{2}}T_{\overline{k},\overline{l},m,n}
+(-)^{l_{2}+m_{2}}T_{k,\overline{l},\overline{m},n}
+(-)^{k_{2}+m_{2}}T_{\overline{k},l,\overline{m},n}\nonumber \\
& &-(-)^{k_{2}+l_{2}+m_{2}}T_{\overline{k},\overline{l},\overline{m},n}
\end {eqnarray}
where the $T_{k,l,m,n,}$ denote the components of the curvature for the torus
(\ref {theformula}). This can be more compactly written by using
$T_{k,l,m,n}=T_{\overline{k},\overline{l},\overline{m},\overline{n}}$ and
$k_{2}+l_{2}+m_{2}+n_{2}=0$ for $R_{k,l,m,n}$ non-zero:
\begin {equation}
R_{e_{k},e_{l},e_{m},e_{n}}=
\frac{1}{2}\sum_{\mu ,\nu ,\rho ,\sigma}\mu ^{k_{2}+1}\nu
 ^{l_{2}+1}\rho ^{m_{2}+1} \sigma
^{n_{2}+1}T_{(\mu k_{1},k_{2}),(\nu l_{1},l_{2}),(\rho m_{1},m_{2})
,(\sigma n_{1},n_{2})}
\end {equation}
where the  Greek variables take the values $1$ or $-1$. In this form the
 symmetries
\begin {equation}
\label {symmetries}
R_{e_{k},e_{l},e_{m},e_{n}}=R_{e_{m},e_{n},e_{k},e_{l}}
=-R_{e_{l},e_{k},e_{m},e_{n}}=-R_{e_{k},e_{l},e_{n},e_{m}}
\end {equation}
are apparent.

\section {The Sectional Curvature}

We now proceed to calculate the sectional curvature for a pair of real
diffeomorphism generators on the Klein bottle. Consider the vectors
\begin {equation}
\label {vectors}
\xi_{k}=(L_{k}+L_{-k})-(-)^{k_{2}}(L_{\overline {k}}+L_{-\overline
{k}})
\end {equation}
By the linearity of the Riemann tensor, we can write
\begin {equation}
\label {reala}
R_{\xi_{k},\xi_{l},\xi_{k},\xi_{l}}
=\sum_{\mu ,\nu ,\rho ,\sigma}
R_{e_{\mu k},e_{\nu l},e_{\rho k},e_{\sigma l}}
\end {equation}
Using the symmetries (\ref {symmetries}), this can be expressed as
\begin {eqnarray}
\label {realb}
R_{\xi_{k},\xi_{l},\xi_{k},\xi_{l}}&=&
4R_{e_{k},e_{l},e_{k},e_{-l}}+4R_{e_{k},e_{l},e_{-k},e_{l}}
+2R_{e_{k},e_{l},e_{k},e_{l}}\nonumber \\
& &+2R_{e_{k},e_{-l},e_{k},e_{-l}}+2R_{e_{k},e_{l},e_{-k},e_{-l}}
+2R_{e_{k},e_{-l},e_{-k},e_{l}}
\end {eqnarray}

Since $T_{k,l,m,n}$ is non-zero only if ${\bf k}+{\bf l}+{\bf m}+{\bf n}=0$,
all terms in $R_{\xi_{k},\xi_{l},\xi_{k},\xi_{l}}$ will vanish unless
 ${\bf k}$ and ${\bf l}$ obey the relations
\begin {eqnarray}
\label {relation1}
\mu k_{1}+\nu l_{1} +\rho k_{1} +\sigma l_{1}&=&0\\
\label {relation2}
\alpha k_{2}+\beta l_{2}+\gamma k_{2}+\delta l_{2}&=&0
\end {eqnarray}
Therefore, the only terms which {\bf always} contribute are those for which
$\mu +\rho =\nu +\sigma =\alpha +\gamma =\beta +\delta =0$. By inspection of
(\ref {realb}), all of these are included in
\begin {displaymath}
2R_{e_{k},e_{l},e_{-k},e_{-l}}+2R_{e_{k},e_{-l},e_{-k},e_{l}}
\end {displaymath}
where, from (\ref {result})
\begin {eqnarray}
R_{e_{k},e_{l},e_{-k},e_{-l}}
&=&T_{k,l,-k,-l}+T_{\overline{k},l,-\overline{k},-l} \\
R_{e_{k},e_{-l},e_{-k},e_{l}}
&=&T_{k,-l,-k,l}+T_{\overline{k},-l,-\overline{k},l}
\end {eqnarray}
and all other terms vanish when ${\bf k}$ and ${\bf l}$ obey no special
relations of the form (\ref {relation1}), (\ref {relation2}). Using (\ref
{theformula}) to calculate these terms, we arrive at
\begin {equation}
R_{\xi_{k},\xi_{l},\xi_{k},\xi_{l}}=
-8\pi ^{2}\left[ \frac{({\bf k}\times {\bf l})^{4}}
{({\bf k}+{\bf l})^{2}} +\frac{(\overline{{\bf k}}\times {\bf l})^{4}}
{(\overline{{\bf k}}+{\bf l})^{2}} +\frac{({\bf k}\times {\bf l})^{4}}
{({\bf k}-{\bf l})^{2}} +\frac{(\overline{{\bf k}}\times {\bf l})^{4}}
{(\overline{{\bf k}}-{\bf l})^{2}}\right]
\end {equation}
Using (\ref {bottlemetric}) to give us the orthogonality relations
$\langle \xi_{k},\xi_{k}\rangle =8\pi ^{2}{\bf k}^{2}$, $\langle \xi_{l}
,\xi_{l}
 \rangle =8\pi ^{2}{\bf l}^{2}$, $\langle \xi_{k},\xi_{l}\rangle =0$, together
 with
(\ref {sectionaldefn}), we express the sectional curvature in a form similar to
that of Arnold \cite {arnold}:
\begin {equation}
\label {generalcurv}
C_{\xi_{k},\xi_{l}}=-\frac{({\bf k}^{2}+{\bf l}^{2})}{16\pi ^{2}}\left[
\sin^{2}\alpha \:\sin ^{2}\beta +\sin^{2}\overline {\alpha}\:
\sin ^{2}\overline {\beta}\right]
\end {equation}
where $\alpha $, $\beta $, $\overline{\alpha}$ and $\overline{\beta}$ are the
angles between ${\bf k}$ and ${\bf l}$, $({\bf k}+{\bf l})$ and $({\bf k}-{\bf
l})$, $\overline{{\bf k}}$ and ${\bf l}$ and $(\overline {{\bf k}}+l)$ and
$(\overline {{\bf k}}-l)$ respectively. For the {\bf general} case, therefore,
the sectional curvature is non-positive.

It remains for us to calculate the sectional curvature for $\xi_{k}$ and
$\xi_{l}$
 such that ${\bf k}$ and ${\bf l}$ have a non-trivial relationship of the form
 (\ref {relation1}), (\ref {relation2}). As an example, we consider the case
$k_{2}=l_{2}$, $k_{1}\neq \pm l_{1}$, where $k_{1},l_{1},k_{2},l_{2}\neq 0$.

In addition to the terms already considered, there is also a contribution from
\begin {displaymath}
2R_{e_{k},e_{-l},e_{k},e_{-l}}
=2T_{\overline {k},-\overline {l},k,-l} +2T_{k,-\overline
{l},\overline {k},-l}
\end {displaymath}
giving the result
\begin {equation}
\label {specialcurv}
C_{\xi_{k},\xi_{l}}=\frac {1}{8\pi ^{2}}\left[
\frac{2({\bf k}\times \overline{{\bf k}})^{2}({\bf l}\times \overline{{\bf
l}})^{2}}{({\bf k}+\overline{{\bf k}})^{2}}
-2\frac{({\bf k}\times {\bf l})^{4}}{({\bf k}-{\bf l})^{2}}
-2\frac{(\overline{{\bf k}}\times {\bf l})^{4}}
{(\overline{{\bf k}}-{\bf l})^{2}}
-\frac{({\bf k}\times {\bf l})^{4}}{({\bf k}+{\bf l})^{2}}
-\frac{(\overline{{\bf k}}\times {\bf l})^{4}}
{(\overline{{\bf k}}+{\bf l})^{2}}\right]
\end {equation}
Calculation in individual cases shows the sign of this quantity to be
indefinite.

Using (\ref {bdefn}), we see that
\begin {equation}
\langle \xi_{k},[\xi_{l},e_{m}]\rangle =\langle e_{m},B(\xi_{k},\xi_{l})\rangle
\end {equation}
and it can be shown that
\begin {equation}
B(\xi_{k},\xi_{k})=0
\end {equation}
so that $\xi_{k}$ generates a stationary flow. As the Klein bottle is locally
Euclidean, it seems that Lukatskii's theorem should apply. In the general case,
there is no disagreement -- however, the special case (\ref {specialcurv}) can
give a positive sectional curvature.

The vectors $\xi_{k}$ also generate diffeomorphisms on the torus, and we
can derive the curvature with respect to these simply by using linearity.
 The result is
\begin {equation}
R_{\xi_{k},\xi_{l},\xi_{k},\xi_{l}}
=\sum_{\mu ,\nu ,\rho ,\sigma}
R_{e_{\mu k},e_{\nu l},e_{\rho k},e_{\sigma l}}
\end {equation}
where
\begin {equation}
R_{e_{k},e_{l},e_{m},e_{n}}=\sum_{\mu ,\nu ,\rho ,\sigma}\mu ^{k_{2}+1}\nu
 ^{l_{2}+1}\rho ^{m_{2}+1} \sigma
^{n_{2}+1}T_{(\mu k_{1},k_{2}),(\nu l_{1},l_{2}),(\rho m_{1},m_{2})
,(\sigma n_{1},n_{2})}
\end {equation}
i.e. the same as for the Klein bottle apart from the absence of the factor
$1/2$. As the form of the scalar product is also the same, the sectional
curvature will have the same sign. Furthermore, from
\begin {equation}
\langle \xi_{k},[\xi_{l},L_{m}]\rangle =\langle L_{m},B(\xi_{k},\xi_{l})\rangle
\end {equation}
we can show that $B(\xi_{k},\xi_{k})=0$, so that the $\xi_{k}$ also generate
stationary flows on the torus. We have therefore found a pair of diffeomorphism
generators on the torus
which disobey Lukatskii's theorem. We now investigate this theorem in detail.

\section {Lukatskii's Theorem}
Lukatskii \cite {lukatskii} gives the following formula for the
(unnormalized)
sectional curvature with respect to the volume-preserving diffeomorphism
generators $u$ and $v$:
\begin {eqnarray}
\label {lukformula}
R_{u,v,u,v}&=&-\frac{1}{4}\left[ h(\chi(u,v),\chi(u,v))+h(\chi(u,u),\chi(v,v))
\right]\\
h(w,x)&\equiv &\langle \hat{w},\hat{x}\rangle \\
\chi(u,v)&\equiv &D_{u}v+D_{v}u
\end {eqnarray}
where $\hat{w}$ is the projection of a vector $w$ on Diff$\cal M$ onto the
space of vectors orthogonal to SDiff$\cal M$, and
\begin {equation}
\label {lukgrad}
D_{u}v=u^{a}\partial_{a}v^{b} \partial_{b}
\end {equation}
for $u=u^{a}\partial_{a}$

We restrict our considerations to the torus.
To calculate the projections, we introduce generators of Diff$\cal M$:
\begin {equation}
w_{k,a}=e^{i{\bf k}\cdot {\bf x}}\partial_{a}
\end {equation}
with the scalar product
\begin {equation}
\langle w_{k,a},w_{l,b}\rangle =4\pi ^{2}\delta_{ab}\delta_{{\bf k}+{\bf l}}
\end {equation}
This is just the extension of (\ref {metricdefn}) to non-volume-preserving
diffeomorphisms.

We denote the vectors orthogonal to the $L_{k}$ by $M_{k}$. We have
\begin {eqnarray}
L_{k}&=&i\epsilon^{ab}k_{b}w_{k,a}\\
M_{k}&=&i\delta^{ab}k_{b}w_{k,a}\\
\langle L_{k},M_{l}\rangle &=&0
\end {eqnarray}
The $L_{k}$ and $M_{k}$ span Diff$\cal M$:
\begin {equation}
\label {shadows}
w_{k,a}=(\epsilon_{ab}L_{k}+\delta_{ab}M_{k})\frac{k^{b}}{i{\bf k}^{2}}
\end {equation}
Using (\ref {lukgrad}), we obtain
\begin {equation}
\chi(L_{k},L_{l})=i({\bf k}\times {\bf l})\epsilon^{ab}(l_{b}-k_{b})w_{k+l,a}
\end {equation}
Projecting onto the space orthogonal to SDiff$\cal M$ by considering the second
term in (\ref{shadows}), we arrive at
\begin {equation}
\label {projection}
\hat{\chi}(L_{k},L_{l})=2\frac{({\bf k}\times {\bf l})^{2}}{({\bf k}+{\bf
l})^{2}}M_{k+l}
\end {equation}

Lukatskii states that if $u$ or $v$ generates a stationary
flow, the second term in (\ref
{lukformula}) vanishes. This is certainly true for the vectors $L_{k}+L_{-k}$
considered by Arnold, and it is easy to show that the use of (\ref
{lukformula}) together with (\ref {projection}) for these vectors
produces his results. However, if we consider
$e_{k}=L_{k}-(-)^{k_{2}}L_{\overline{k}}$ on the torus then we obtain, by
linearity:
\begin {equation}
\hat{\chi}(e_{k},e_{l})=2\frac{({\bf k}\times {\bf l})^{2}}{({\bf k}+{\bf
l})^{2}}\left[ M_{k+l}+(-)^{k_{2}+l_{2}}M_{\overline{k}+\overline{l}}\right]
-2\frac{(\overline{{\bf k}}\times {\bf l})^{2}}{(\overline{{\bf k}}+{\bf
l})^{2}}\left[ (-)^{k_{2}}M_{\overline{k}+l}+(-)^{l_{2}}M_{k+\overline{l}}
\right]
\end {equation}
so that
\begin {equation}
\hat{\chi}(e_{k},e_{k})=-4(-)^{k_{2}}\frac{({\bf k}\times \overline{{\bf
k}})^{2}}{({\bf k}+\overline{{\bf k}})^{2}}M_{k+\overline{k}}
\end {equation}
giving
\begin {equation}
\langle \hat{\chi}(e_{k},e_{k}),\hat{\chi}(e_{l},e_{l})\rangle
=64\pi^{2}\frac{({\bf k}\times \overline{{\bf k}})^{2}({\bf l}\times
\overline{{\bf l}})^{2}}{({\bf k}+\overline{{\bf k}})^{2}}
\delta_{{\bf k}+{\bf l}+\overline{{\bf k}}+\overline{{\bf l}}}
\end {equation}
We see that this term is non-zero if ${\bf k}+{\bf l}+\overline{\bf k}
+\overline{\bf l}=0$, ie: if $k_{2}+l_{2}=0$. As $B(e_{k},e_{k})=0$, the
 $e_{k}$ generate
stationary flows, and we deduce that Lukatskii's statement is incorrect. Note,
however, that for most ${\bf k}$ and ${\bf l}$, the term vanishes, and
Lukatskii's theorem holds.

For $k_{2}+l_{2}=0$, $k_{1}\neq \pm l_{1}$ and
 $k_{1},l_{1},k_{2},l_{2}\neq 0$,
the other term in (\ref {lukformula}) is
\begin {equation}
\langle \hat{\chi}(e_{k},e_{l}),\hat{\chi}(e_{k},e_{l})\rangle =
32\pi ^{2}\left[\frac{({\bf k}\times {\bf l})^{4}}{({\bf k}+{\bf l})^{2}}
+\frac{(\overline{{\bf k}}\times{\bf l})^{4}}{(\overline{{\bf k}}+{\bf
l})^{2}}\right]
\end {equation}
 We suspect further that there is a misprint in \cite {lukatskii},
 and that (\ref {lukformula}) should be
\begin {equation}
\label {signchange}
R_{u,v,u,v}=-\frac{1}{4}\left[ h(\chi(u,v),\chi(u,v))-h(\chi(u,u),\chi(v,v))
\right]
\end {equation}
Then
\begin {equation}
R_{e_{k},e_{l},e_{k},e_{l}}=8\pi ^{2}\left[
2\frac{({\bf k}\times \overline{{\bf k}})^{2}({\bf l}\times \overline{{\bf
l}})^{2}}{({\bf k}+\overline{{\bf k}})^{2}}
-\frac{({\bf k}\times {\bf l})^{4}}{({\bf k}+{\bf l})^{2}}
-\frac{(\overline{{\bf k}}\times {\bf l})^{4}}{(\overline{{\bf k}}+{\bf
l})^{2}}\right]
\end {equation}
which agrees with the result derived by using the linearity of the Riemann
tensor.
 Similarly, it can be shown that (\ref {signchange}) produces the
correct results for a general
 linear combination of the $L_{k}$. In particular, it
can be used to derive the expression (\ref {specialcurv}) for the real flows
$\xi_{k}$ and $\xi_{l}$, where $k_{2}=l_{2}$.

Using the first term in (\ref {shadows}), we see that
\begin {equation}
P[D_{L_{k}}L_{l}]=\frac {({\bf k}\times {\bf l}){\bf l}\cdot ({\bf k}+{\bf l})}
{({\bf k}+{\bf l})^{2}}L_{k+l}
\end {equation}
where P[ ] denotes the projection onto the Lie algebra of
SDiff$\cal M$. This gives
\begin {equation}
P[D_{L_{k}}L_{l}]=\nabla _{L_{k}}L_{l}
\end {equation}
By linearity, this holds for all vectors. In fact, Nakamura et al \cite
{nakamura} use $D_{u}v$ as the definition of $\nabla_{u}v$. The calculations
are unaffected since only the projection onto SDiff$\cal M$ is used.
(\ref {geo}) gives us
\begin {equation}
\frac {\partial v}{\partial t}=-P[D_{v}v]
\end {equation}
If $D_{v}v=0$, then $v$ certainly generates a stationary flow, and Lukatskii's
theorem holds, since $\hat{\chi}(v,v)=0$. However, in general, we only require
that $P[D_{v}v]=0$, for which $\hat{\chi}(v,v)$ does not necessarily vanish.
Lukatskii's theorem therefore only works for a subclass of stationary flows.

\section {The Projective Plane}

The projective plane can be constructed from a $2\pi \times 2\pi$
plane by making two reflections:
\begin {eqnarray}
(x,y)&=&(x+\pi ,\pi -y)\\
(x,y)&=&(\pi -x,y+\pi )
\end {eqnarray}
As stated in \cite {pope},
generators which are invariant under these identifications have the form
\begin {equation}
\label {progenerator}
f_{k}=(L_{k}+L_{-k})-(-)^{k_{1}+k_{2}}(L_{\overline{k}}+L_{-\overline{k}})
\end {equation}

As the calculation is similar to that for the Klein bottle, we do not go into
detail. The components of the
Riemann tensor for the group of diffeomorphisms on the
projective plane are
\begin {eqnarray}
R_{f_{k},f_{l},f_{m},f_{n}}&=&\frac{1}{4}
\sum_{\alpha ,\beta ,\gamma ,\delta}\sum_{\mu ,\nu ,\rho ,\sigma}[
(\alpha \mu)^{k_{1}+k_{2}+1}(\beta \nu)^{l_{1}+l_{2}+1}(\gamma
\rho)^{m_{1}+m_{2}+1}(\delta \sigma)^{n_{1}+n_{2}+1}\nonumber \\
& &\:\:\:\:\:\:\:\:\:\:\:\:\:\:\:\:\:\:\:\:\:\:\:\:\:\:\:T_{(\mu k_{1},\alpha
k_{2}),(\nu l_{1},\beta l_{2}),(\rho m_{1},\gamma m_{2}),(\sigma n_{1},\delta
n_{2})}]
\end {eqnarray}
The sectional curvature with respect to $f_{k}$ and $f_{l}$ is given exactly by
(\ref {generalcurv}) in the general case. In special cases of the form (\ref
{relation1}) and (\ref {relation2}) the sign of the sectional curvature is
indefinite.

Alternatively, a curved version of the projective plane can be formed by
identifying antipodal points on a sphere, ie:
\begin {equation}
\label {proident}
(\theta,\phi)=(\pi-\theta,\phi+\pi)
\end {equation}
On this manifold, the spherical harmonics $Y_{lm}(\theta,\phi)$
 are a complete set of basis functions.
 The diffeomorphism generators can therefore be constructed from
\begin {equation}
L_{lm}=\frac{1}{\sin\theta}\epsilon^{ab}\partial_{b}Y_{lm}\partial_{a}
\end {equation}
These are invariant under (\ref {proident}) if and only if $l$ is odd.

By symmetry, we have
\begin {equation}
\langle L_{l_{1}m_{1}},L_{l_{2}m_{2}}\rangle
_{projective\:\:plane}=
\frac{1}{2}\langle L_{l_{1}m_{1}},L_{l_{2}m_{2}}\rangle_{sphere}
\end {equation}
Using this together with
the expressions for the Lie bracket and covariant derivative derived by
Arakelyan and Savvidy \cite {arakelyan} for the sphere:
\begin {eqnarray}
[L_{l_{1}m_{1}},L_{l_{2}m_{2}}]&=&\sum_{l_{3},m_{3}}
G_{l_{1}m_{1}l_{2}m_{2}}^{l_{3}m_{3}}
L_{l_{3}m_{3}}\\
\nabla_{L_{l_{1}m_{1}}}L_{l_{2}m_{2}}&=&\sum _{l_{3},m_{3}}
\Gamma_{l_{1}m_{1}l_{2}m_{2}}^{l_{3}m_{3}}
L_{l_{3}m_{3}}
\end {eqnarray}
\begin {equation}
G_{l_{1}m_{1}l_{2}m_{2}}^{l_{3}m_{3}}=
\Gamma_{l_{1}m_{1}l_{2}m_{2}}^{l_{3}m_{3}}=0,\:\:l_{1}+l_{2}+l_{3}\:\:
\mbox{even}
\end {equation}
we see that we can calculate the curvature of
SDiff(projective plane) simply by considering the curvature with respect to the
 $L_{lm}$, where $l$ is odd, on the sphere. For an expression for the
curvature, we refer to \cite {arakelyan}.

\end {document}